\documentclass[bdcc,communication,accept,oneauthor,pdftex,10pt,a4paper]{mdpi_arxiv}

\firstpage{1}
\articlenumber{21}
\pubvolume{3}
\issuenum{2}
\pubyear{2019}
\copyrightyear{2019}
\history{Received: 28 February 2019; Accepted: 29 March 2019; Published: 5 April 2019}

\usepackage{changepage}
\usepackage{environ}
\NewEnviron{myequation}{%
\begin{equation}
\scalebox{0.95}{$\BODY$}
\end{equation}}

\theoremstyle{mdpi}
\newcounter{thm}
\theoremstyle{mdpidefinition}
\newtheorem*{Model}{Model} 
\newtheorem{FailureMode}[thm]{Failure Mode}

\Title{Multiparty Dynamics and Failure Modes for Machine Learning and Artificial Intelligence}

\Author{David Manheim}
\AuthorNames{David Manheim}

\abstract{An important challenge for safety in machine learning and artificial intelligence systems is a~set of related failures involving specification gaming, reward hacking, fragility to distributional shifts, and Goodhart's or Campbell's law. This  paper presents additional failure modes for interactions within multi-agent systems that are closely related. These multi-agent failure modes are more complex, more problematic, and less well understood than the single-agent case, and are also already occurring, largely unnoticed. After motivating the discussion with examples from poker-playing artificial intelligence (AI), the paper  explains why these failure modes are in some senses unavoidable. Following this, the paper categorizes failure modes, provides definitions, and cites examples for each of the modes: accidental steering, coordination failures, adversarial misalignment, input spoofing and filtering, and goal co-option or direct hacking. The paper then discusses how extant literature on multi-agent AI fails to address these failure modes, and identifies work which may be useful for the mitigation of these failure modes.}

\keyword{multi-agent systems; specification gaming; artificial intelligence safety; Goodhart's Law}

\MSC{91E45; 91A06}
\JEL{C79; D74}

\begin{document}




\section{Background, Motivation and Contribution}

When complex systems are optimized by a single agent, the representation of the system and of the goal used for optimization often leads to failures that can be surprising to the agent's designers. These failure modes go by a variety of names, Amodei and Clark called them faulty reward functions~\cite{Clark2016} but similar failures have been referred to as Goodhart's law~\cite{Goodhart1975,Manheim2018b}, Campbell's law~\cite{Campbell1979}, distributional shift~\cite{Amodei2016}, strategic behavior~\cite{kleinberg2018classifiers}, reward hacking~\cite{Bostrom2017}, Proxyeconomics~\cite{braganza2018proxyeconomics},  and other terms.

Examples of these failures in the single-agent case are shown by Victoria Krakovna's extensive list of concrete examples of ``generating a solution that literally satisfies the stated objective but fails to solve the problem according to the human designer’s intent.'' \cite{Krakovna2018} Liu et al. suggest that ``a complex activity can often be performed in several different ways,'' \cite{LiuEtAl2016} but not all these ways should be considered valid. To understand why, Krakovna' s list includes examples of ``achieving a goal'' by finding and exploiting bugs in a simulation engine to achieve goals~\cite{Cheney2014, Figueras2015, Lehman2018}; by physical manipulation of objects in unanticipated ways, such as moving a table instead of the item on the table~\cite{Chopra2018}, or flipping instead of lifting a block~\cite{Popov2017}; and even by exploiting the problem structure or evaluation, such as returning an empty list as being sorted~\cite{Weimer2013}, or deleting the file containing the target output~\cite{Weimer2013}.

\subsection{Motivation}

This forms only a part of the broader set of concerns in AI safety,~\cite{Sandberg2001, Yudkowsky2011, Amodei2016, WorleyIII2018}, but the failure modes are the focus of a significant body of work in AI safety discussed later in the paper. However, as the systems become more capable and more widely used, Danzig and others have noted that this will ``increase rather than reduce collateral risks of loss of control.'' \cite{Danzig2018} The speed of such systems is almost certainly beyond the point of feasible human control, and as they become more complex, the systems are also individually likely to fail in ways that are harder to understand.

While some progress has been made in the single-agent case, the systems have continued to become more capable, corporations, governments, and other actors have developed and deployed machine learning systems. These are not only largely autonomous, but also interact with each other. This allows a new set of failures, and these are not yet a focus of safety-focused research---but they are~critical.

\subsection{Contribution}

The analogues of the earlier-mentioned classes of failure for multi-agent systems are more complex, potentially harder to mitigate, and unfortunately not the subject of a significant focus among AI safety researchers. In this paper, we introduce a classification of failures that are not yet well-addressed in the literature involving multiple agents. These failures can occur even when system designers do not intend to build conflicting AI or ML systems. The current paper contributes to the literature by outlining how and why these multi-agent failures can occur, and providing an overview of approaches that could be developed for mitigating them. In doing so, the paper will hopefully help spur system designers to explicitly consider these failure modes in designing systems, and urge caution on the part of policymakers.

As a secondary contribution, the link between ongoing work on AI safety and potential work mitigating these multi-agent failures incidentally answers an objection raised by AI risk skeptics that AI safety is ''not worth current attention'' and that the issues are ``premature to worry about'' \cite{Baum2018}. This paper instead shows how failures due to multi-agent dynamics are critical in the present, as ML and superhuman narrow AI is being widely deployed, even given the (valid) arguments put forward by Yudkowsky~\cite{Yudkowsky2013} and Bostrom~\cite{Bostrom2017} for why a singleton AI is a more important source of existential risk.

\subsection{Extending Single-Agent Optimization Failures}
Systems which are optimized using an imperfect system model have several important failure modes categorized in work by Manheim and Garrabrant~\cite{Manheim2018b}. First, imperfect correlates of the goal will be less correlated in the tails of the distribution, as discussed by Lewis~\cite{Lewis2014}. Heavily optimized systems will end up in those regions, and even well-designed metrics do not account for every possible source of variance. Second, there are several context failures~\cite{Yudkowsky2016b}, where the optimization is well behaved in the training set (``ancestral environment'') but fails as optimization pressure is applied. For example, it may drift towards an ``edge instantiation'' where the system may optimize all the variables that relate to the true goal, but further gain on the metric is found by unexpected means. Alternatively, the optimizer may properly obey constraints in the initial stage, but find some ``nearest unblocked strategy'' \cite{Yudkowsky2016b} allowing it to circumvent designed limits when given more optimization power. These can all occur in single-agent scenarios.

The types of failure in multi-agent systems presented in this paper can be related to Manheim and Garrabrant's classification of single-agent metric optimization failures . The four single-agent overoptimization failure modes outlined there are:
\begin{itemize}[leftmargin=*,labelsep=4mm]
\item Tails Fall Apart, or Regressional inaccuracy, where the relationship between the modeled goal and the true goal is inexact due to noise (for example, measurement error,) so that the bias grows as the system is optimized.
\item Extremal Model Insufficiency, where the approximate model omits factors which dominate the system's behavior after optimization.
\item Extremal Regime Change, where the model does not include a regime change that occurs under certain (unobserved) conditions that optimization creates.
\item Causal Model Failure, where the agent's actions are based on a model which incorrectly represents causal relationships, and the optimization involves interventions that break the causal structure the model implicitly relies on.
\end{itemize}

Despite the completeness of the above categorization, the way in which these failures occur can differ greatly even when only a single agent is present. In a multi-agent scenario, agents can stumble into or intentionally exploit model overoptimization failures in even more complex ways. Despite this complexity, the different multi-agent failure modes can be understood based on understanding the way in which the implicit or explicit system models used by agents fail.

\subsection{Defining Multi-Agent Failures}
In this paper, a multi-agent optimization failure is when one (or more) of the agents which can achieve positive outcomes in some scenarios exhibits behaviors that negatively affect its own outcome due to the actions of one or more agents other than itself. This occurs either when the objective function of the agent no longer aligns with the goal, as occurs in the Regressional and both Extremal cases, or~when the learned relationship between action(s), the metric(s), and the goal have changed, as in the Causal failure case.

This definition does not require the failure to be due to malicious behavior on the part of any agent, nor does it forbid it. Note also that the definition does not require failure of the system, as in Behzadan and Munir's categorization of adversarial attacks~\cite{Behzadan2017}, nor does it make any assumptions about type of the agents, such as the type of learning or optimization system used. (The multi-agent cases implicitly preclude agents from being either strongly boxed, as Drexler proposed~\cite{Drexler1986}, or oracular, as discussed by Armstrong~\cite{Armstrong2012}.)

\section{Multi-Agent Failures: Context and Categorization}

Several relatively straightforward failure modes involving interactions between an agent and a~regulator were referred to in Manheim and Garrabrant as adversarial Goodhart~\cite{Manheim2018b}. These occur where one AI system opportunistically alters or optimizes the system and uses the expected optimization of a~different victim agent to hijack the overall system. For example, ``smart market'' electrical grids use systems that optimize producer actions and prices with a linear optimization system using known criteria. If power lines or power plants have strategically planned maintenance schedules, an owner can manipulate the resulting prices to its own advantage, as occurred (legally) in the case of Enron~\cite{Mulligan2002LATimes}. This is possible because the manipulator can plan in the presence of a known optimization regime.

This class of manipulation by an agent frustrating a regulator's goals is an important case, but~more complex dynamics can also exist, and Manheim and Garrabrant noted that there are ``clearly further dynamics worth exploring.'' \cite{Manheim2018b} This involves not only multiple heterogenous agents, which Kleinberg and Raghavan suggest an avenue for investigating, but also interaction between those agents~\cite{kleinberg2018classifiers}. An example of a well-understood multi-agent system, the game of poker, allows clarification of why the complexity is far greater in the interaction case.

\subsection{Texas Hold'em and the Complexity of Multi-Agent Dynamics}
In many-agent systems, simple interactions can become complex adaptive systems due to agent behavior, as the game of poker shows. Solutions to simplified models of two-player poker predate game theory as a field~\cite{Borel1938}, and for simplified variants, two-player draw poker has a fairly simple optimal strategy~\cite{Kuhn1950}. These early, manually computed solutions were made possible both by limiting the complexity of the cards, and more importantly by limiting interaction to a single bet size, with no raising or interaction between the players. In the more general case of heads-up limit Texas Hold'em, significantly more work was needed, given the multiplicity of card combinations, the existence of hidden information, and player interaction, but this multi-stage interactive game is ``now essentially weakly solved'' \cite{Bowling2015}. Still, this game involves only two players. In the no-limit version of the game, Brown and Sandholm recently unveiled superhuman AI~\cite{Browneaao1733}, which restricts the game to ``Heads' Up'' poker, which involves only two players per game, and still falls far short of a full solution to the game.

The complex adaptive nature of multi-agent systems means that each agent needs model not only model the system itself, but also the actions of the other player(s). The multiplicity of potential outcomes, betting strategies, and different outcomes becomes rapidly infeasible to represent other than heuristically. In limit Texas Hold'em poker, for example, the number of card combinations is immense, but the branching possibilities for betting is the more difficult challenge. In a no-betting game of Hold'em with P players, there are $52!/((52-2P-5)!\cdot 2P \cdot 5!)$ possible situations. This is $2.8\cdot10^{12}$ hands in the two-player case, $3.3\cdot10^{15}$ in the three-player case, and growing by a similar factor when expanded to the four-, five-, or six-player case. The probability of winning is the probability that the five cards on the table plus two unknown other cards from the deck are a better hand than any that another player holds. In Texas Hold'em, there are four betting stages, one after each stage of cards is revealed. Billings~et~al. use a reduced complexity game (limiting betting to three rounds per stage) and find a complexity of $O(10^{18})$ in the two-hand case~\cite{Billings2003}. That means the two-player, three-round game complexity is comparable in size to a no-betting four-player game, with $4.1\cdot10^{18}$ card combinations~possible.

Unlike a no-betting game, however, a player must consider much more than the simple probability that the hand held is better than those held by other players. That calculation is unmodified during the additional branching due to player choices. The somewhat more difficult issue is that the additional branching requires Bayesian updates to estimate the probable distribution of hand strengths held by other players based on their decisions, which significantly increases the complexity of solving the game. The most critical challenge, however, is that each player bets based on the additional information provided by not only the hidden information provided by their cards, but also based on the betting behavior of other players. Opponent(s) make betting decisions based on non-public information (in~Texas Hold'em, hole cards) and strategy for betting requires a meta-update taking advantage of the information the other player reveals by betting. The players must also update based on potential strategic betting by other players, which occurs when a player bets in a way calculated to deceive. To~deal with this, poker players need to model not just the cards, but also the strategic decisions of other players. This complex model of strategic decisions must be re-run for all the possible combinations at each decision point to arrive at a conclusion about what other players are doing. Even after this is complete, an advanced poker player, or an effective AI, must then decide not just how likely they are to win, but also how to play strategically, optimizing based on how other players will react to the different choices available.

Behaviors such as bluffing and slow play are based on these dynamics, which become much more complex as the number of rounds of betting and the number of players increases. For example, slow play involves underbidding compared to the strength of your hand. This requires that the players will later be able to raise the stakes, and allows a player to lure others into committing additional money. The complexity of the required modeling of other agents' decision processes grows as a function of the number of choices and stages at which each agent makes a decision. This type of complexity is common in multi-agent systems. In general, however, the problem is much broader in scope than what can be illustrated by a rigidly structured game such as poker.

\subsection{Limited Complexity Models versus the Real World}
In machine learning systems, the underlying system is approximated by implicitly or explicitly learning a multidimensional transformation between inputs and outputs. This transformation approximates a combination of the relationships between inputs and the underlying system, and~between the system state and the outputs. The complexity of the model learned is limited by the computational complexity of the underlying structure, and while the number of possible states for the input is large, it is typically dwarfed by the number of possible states of the system.

The critical feature of machine learning that allows such systems to be successful is that most relationships can be approximated without inspecting every available state. (All models simplify the systems they represent.) The implicit simplification done by machine learning is often quite impressive, picking up on clues present in the input that humans might not notice, but it comes at the cost of having difficult to understand and  difficult to interpret implicit models of the system.

Any intelligence, whether machine learning-based, human, or AI, requires similar implicit simplification, since the branching complexity of even a relatively simple game such as Go dwarfs the number of atoms in the universe. Because even moderately complex systems cannot be fully represented, as discussed by Soares~\cite{Soares2014}, the types of optimization failures discussed above are inevitable. The contrapositive to Conant and Ashby's theorem~\cite{Conant1970} is that if a system is more complex than the model, any attempt to control the system will be imperfect. Learning, whether human or machine, builds approximate models based on observations, or input data. This implies that the behavior of the approximation in regions far from those covered by the training data is more likely to markedly differ from reality. The more systems change over time, the more difficult prediction becomes---and the more optimization is performed on a system, the more it will change. Worsening this problem, the learning that occurs in ML systems fails to account for the embedded agency issues discussed by Demski and Garrabrant~\cite{Demski2019}, and interaction between agents with implicit models of each other and themselves amplifies many of these concerns.

\subsection{Failure modes}

Because an essential part of multi-agent dynamic system modeling is opponent modeling, the opponent models are a central part of any machine learning model. These opponent models may be implicit in the overall model, or they may be explicitly represented, but they are still models that are approximate. In many cases, opponent behavior is ignored---by implicitly simplifying other agent behavior to noise, or by assuming no adversarial agents exist. Because these models are imperfect, they will be vulnerable to overoptimization failures discussed above.

The list below is conceptually complete, but limited in at least three ways. First, examples given in this list primarily discuss failures that occur between two parties, such as a malicious actor and a victim, or failures induced by multiple individually benign agents. This would exclude strategies where agents manipulate others indirectly, or those where coordinated interaction between agents is used to manipulate the system. It is possible that when more agents are involved, more specific classes of failure will be relevant.

Second, the below list does not include how other factors can compound metric failures. These are critical, but may involve overoptimization, or multiple-agent interaction, only indirectly. For~example, O'Neil discusses a class of failure involving the interaction between the system, the inputs, and validation of outputs~\cite{ONeil2016}. These failures occur when a system's metrics are validated in part based on outputs it contributes towards. For example, a system predicting greater crime rates in areas with high minority concentrations leads to more police presence, which in turn leads to a higher rate of crime found. This higher rate of crime in those areas is used to train the model, which leads it to reinforce the earlier unjustified assumption. Such cases are both likely to occur, and especially hard to recognize, when the interaction between multiple systems is complex, and it is unclear whether the system's effects are due in part to its own actions (This class of failure seems particularly likely in systems that are trained via ''self-play,'' where failures in the model of the system get reinforced by incorrect feedback on the basis of the models, which is also a case of model insufficiency failure.). 

Third and finally, the failure modes exclude cases that do not directly involve metric overoptimizations, such as systems learning unacceptable behavior implicitly due to training data that contains unanticipated biases, or failing to attempt to optimize for social preferences such as fairness. These are again important, but they are more basic failures of system design.

With those caveats, we propose the following classes of multi-agent overoptimization failures. For~each, a general definition is provided, followed by one or more toy models that demonstrate the failure mode.  Each agent attempts to achieve their goal by optimizing for the metric, but the optimization is performed by different agents without any explicit coordination or a priori knowledge about the other agents. The specifics of the strategies that can be constructed and the structure of the system can be arbitrarily complex, but as explored below, the ways in which these models fail can still be understood~generally.

These models are deliberately simplified, but where possible, real-world examples of the failures exhibited in the model are suggested. These examples come from both human systems where parallel dynamics exist, and examples of the failures in extent systems with automated agents. In the toy models, $M_i$ and $G_i$ stands for the metric and goal, respectively, for agent $i$. The metric is an imperfect proxy for the goal, and will typically be defined in relation to a goal. (The goal itself is often left unspecified, since the model applies to arbitrary systems and agent goals.) In some cases, the failure is non-adversarial, but where relevant, there is a victim agent $V$ and an opponent agent $O$ that attempts to exploit it. Please note that the failures can be shown with examples formulated with game-theoretic notation, but doing so requires more complex specifications of the system and interactions than is possible using the below characterization of the agent goals and the systems.

\begin{FailureMode} \textbf{Accidental Steering} is when multiple agents alter the systems in ways not anticipated by at least one agent, creating one of the above-mentioned single-party overoptimization failures.\end{FailureMode}

\begin{Remark} This failure mode manifests similarly to the single-agent case and differs only in that agents do not anticipate the actions of other agents. When agents have closely related goals, even if those goals are aligned, it~can exacerbate the types of failures that occur in single-agent cases.

Because the failing agent alone does not (or cannot) trigger the failure, this differs from the single-agent case. The distributional shift can occur due to a combination of actors' otherwise potentially positive influences by either putting the system in an extremal state where the previously learned relationship decays, or triggering a regime change where previously beneficial actions are harmful.\end{Remark}
\begin{Model}\textbf{1.1---Group Overoptimization.}
A set of agents each have goals which affect the system in related ways, and metric-goal relationship changes in the extremal region where x>a. As noted above, $M_i$ and $G_i$ stands for the metric and goal, respectively, for agent $i$. This extremal region is one where single-agent failure modes will occur for some or all agents. Each agent $i$ can influence the metric by an amount $\alpha_i$, where $\sum \alpha_i > a$, but~$\forall \alpha_i < a$. In the extremal subspace where $ M_i > a$, the metric reverses direction, making further optimization of the metric harm the agent's goal.
\begin{equation}
M_i =
\begin{cases}
G_i, & \text{where  } M_i <= a \\
M_i(a)-G_i, & \text{where  } M_i > a \\
\end{cases}
\end{equation}
\end{Model}
\begin{Remark} In the presence of multiple agents without coordination, manipulation of factors not already being manipulated by other agents is likely to be easier and more rewarding, potentially leading to inadvertent steering due to model inadequacy, as discussed in Manheim and Garrabrant's categorization of single-agent cases~\cite{Manheim2018b}. As~shown there overoptimization can lead to perverse outcomes, and the failing agent(s) can hurt both their own goals, and in similar ways, can lead to negative impacts on the goals of other agents.\end{Remark}
\begin{Model} \textbf{ 1.2---Catastrophic Threshold Failure.}
\begin{myequation}
M_i = x_i \hspace*{3.5cm}
\begin{split}
G_i =     \begin{cases}
a + (\sum_{\forall i}x_i)  & \text{where } \sum_{\forall i}x_i <= T \\
a - (\sum_{\forall i}x_i) & \text{where } \sum_{\forall i}x_i > T
\end{cases}
\end{split}
\end{myequation}
Each agent manipulates their own variable, unaware of the overall impact. Even though the agents are collaborating, because they cannot see other agents' variables, there is no obvious way to limit the combined impact on the system to stay below the catastrophic threshold $T$. Because each agent is exploring a different variable, they each are potentially optimizing different parts of the system.
\end{Model}
\begin{Remark} This type of catastrophic threshold is commonly discussed in relations to complex adaptive systems, but can occur even in systems where the catastrophic threshold is simple. The case discussed by Michael Eisen involves pricing on Amazon was due to a pair of deterministic linear pricing-setting bots interacting to set the price of an otherwise unremarkable biology book at tens of millions of dollars, showing that runaway dynamics are possible even in the simplest cases~\cite{Eisen2011}. This phenomenon is also expected whenever exceeding some constraint breaks the system, and such constraints are often not identified until a failure occurs.\end{Remark}
\begin{Example} This type of coordination failure can occur in situations such as overfishing across multiple regions, where each group catches local fish, which they can see, but at a given threshold across regions the fish population collapses, and recovery is very slow. (In this case, the groups typically are selfish rather than collaborating, making the dynamics even more extreme.)\end{Example}
\begin{Example} Smaldino and McElreath~\cite{Smaldino160384} shows this failure mode specifically occurring with statistical methodology in academia, where academics find novel ways to degrade statistical rigor. The more general ``Mutable Practices'' model presented by Braganza~\cite{braganza2018proxyeconomics}, based on part on Smaldino and McElreath, has each agent attempting to both outperform the other agents on a metric as well as fulfill a shared societal goal, allows agents to evolve and find new strategies that combine to subvert a societal goal.\end{Example}

\begin{FailureMode} \textbf{Coordination Failure} occurs when multiple agents clash despite having potentially compatible goals.\end{FailureMode}
\vspace{-15pt}\begin{Remark} Coordination is an inherently difficult task, and can in general be considered impossible~\cite{Gibbard1973}. In~practice, coordination is especially difficult when the goals of other agents are incompletely known or not fully understood. Coordination failures such as Yudkowsky's Inadequate equilibria are stable, and coordination to escape from such an equilibrium can be problematic even when agents share goals~\cite{Yudkowsky2017}.\end{Remark}
\begin{Model}{\textbf{2.1---Unintended Resource Contention.} }
A fixed resource $R$ is split between uses $R^n$ by different agents. Each agent has limited funds $f_i$, and $R_i$ is allocated to agent i for exploitation in proportion to their bid for the resources $c_{R_i}$. The agents choose amounts to spend on acquiring resources, and then choose amounts $s_{n_i}$ to exploit each resource, resulting in utility $U(s_n,R_n)$. The agent goals are based on the overall exploitation of the resources by all agents.
\begin{equation}
\begin{split}
R_i = \frac{c_{R_i}}{\sum_{\forall i}^{}c_{R_i}} \\
G_i = \sum_{\forall i}U_{i,n}(s_{n_i}, R_n)
\end{split}
\end{equation}
In this case, we see that conflicting instrumental goals that neither side anticipates will cause wasted funds due to contention. The more funds spent on resource capture, which is zero-sum, the less remaining for exploitation, which can be positive-sum. Above nominal spending on resources to capture them from aligned competitor-agents will reduce funds available for exploitation of those resources, even though less resource contention would benefit all agents.
\begin{Remark} Preferences and gains from different uses can be homogeneous, so that all agents have no marginal gain from affecting the allocation, funds will be wasted on resource contention. More generally, heterogeneous preferences can lead to contention to control the allocation, with sub-optimal individual outcomes, and~heterogeneous abilities can lead to less-capable agents harming their goals by capturing then ineffectively exploiting resources.\end{Remark}
\end{Model}
\begin{Example} Different forms of scientific research benefit different goals differently. Even if spending in every area benefits everyone, a fixed pool of resources implies that with different preferences, contention between projects with different positive impacts will occur. To the extent that effort must be directed towards grant-seeking instead of scientific work, the resources available for the projects themselves are reduced, sometimes enough to cause a~net~loss.\end{Example}
\begin{Remark} Coordination limiting overuse of public goods is a major area of research in economics. Ostrom explains how such coordination is only possible when conflicts are anticipated or noticed and where a reliable mechanism can be devised~\cite{Ostrom1990}.\end{Remark}
\begin{Model}{\textbf{2.2---Unnecessary Resource Contention. }} As above, but each agent has an identical reward function of $f_{i,n}$.
Even though all goals are shared, a lack of coordination in the above case leads to overspending, as~shown in simple systems and for specified algebraic objective functions in the context of welfare economics. This literature shows many methods for how gains are possible, and in the simplest examples this occurs when agents coordinate to minimize overall spending on resource acquisition.\end{Model}
\begin{Remark} Coordination mechanisms themselves can be exploited by agents. The field of algorithmic game theory has several results for why this is only sometimes possible, and how building mechanisms to avoid such exploitation is possible~\cite{nisan2007algorithmic}.\end{Remark}
\begin{FailureMode} \textbf{Adversarial optimization} can occur when a victim agent has an incomplete model of how an opponent can influence the system. The opponent's model of the victim allows it to intentionally select for cases where the victim's model performs poorly and/or promotes the opponent's goal~\cite{Manheim2018b}.\end{FailureMode}
\vspace{-15pt}\begin{Model}{\textbf{3.1---Adversarial Goal Poisoning.}}
\begin{equation}
\begin{aligned}[c]
G_V&= x\\
G_O&=-x
\end{aligned}
\hspace{20px}
\begin{aligned}[c]
M_V&=X: X \sim normal(x,\sigma^2(y))\\
M_O&=(X,y)
\end{aligned}
\end{equation}
In this case, the Opponent $O$ can see the metric for the victim, and can select for cases where y is large and X is small, so that $V$ chooses maximal values of X, to the marginal benefit of $O$. \end{Model}
\begin{Example} A victim's model can be learned by ``Stealing'' models using techniques such as those explored by Tram{\`e}r et al.~\cite{tramer2016stealing}. In such a case, the information gained can be used for model evasion and other attacks mentioned there.\end{Example}
\begin{Example} Chess and other game engines may adaptively learn and choose openings or strategies for which the victim is weakest.\end{Example}
\begin{Example} Sophisticated financial actors can make trades to dupe victims into buying or selling an asset (``Momentum Ignition'') in order to exploit the resulting price changes~\cite{shorter2014high}, leading to a failure of the exploited agent due to an actual change in the system which it misinterprets.\end{Example}
\begin{Remark} The probability of exploitable reward functions increases with the complexity of the system the agents manipulate~\cite{Amodei2016}, and the simplicity of the agent and their reward function. The potential for exploitation by other agents seems to follow the same pattern, where simple agents will be manipulated by agents with more accurate opponent models.\end{Remark}
\begin{Model}{\textbf{3.2---Adversarial Optimization Theft.}} An attacker can discover exploitable quirks in the goal function to make the victim agent optimize for a new goal, as in Manheim and Garrabrant's Campbell's law example, slightly adapted here~\cite{Manheim2018b}.
\begin{equation} \begin{split}  M_V &= G_V + X \\
M_O &= G_O\cdot X
\end{split}  \end{equation}
$O$ selects $M_O$ after seeing $V$'s choice of metric.
In this case, we can assume the opponent chooses a metric to maximize based on the system and the victim's goal, which is known to the attacker. The opponent can choose their $M_O$ so that the victim's later selection then induces a relationship between $X$ and the opponent goal, especially at the extremes. Here, the opponent selects such that even weak selection on $M_O$ hijacks the victim's selection on $M_V$ to achieve their goal, because states where $M_V$ is high have changed. In the example given, if~$X \sim normal(\mu, \sigma^2)$, the correlation between $G_O$ and $M_O$ is zero over the full set of states, but becomes positive on the subspace selected by the victim. (Please note that the opponent choice of metric is not itself a~useful proxy for their goal absent the victim's actions---it is a purely parasitic choice.)\end{Model}
\begin{FailureMode} \textbf{Input spoofing and filtering}---Filtered evidence can be provided, or false evidence can be manufactured and put into the training data stream of a victim agent.\end{FailureMode}

\begin{Model}{\textbf{4.1---Input Spoofing.}}
Victim agent receives public data $D(x_i|t)$ about the present world-state, and builds a model to choose actions which return rewards $f(x|t)$. The opponent can generate events $x_i$ to poison the victim's learned model.
\end{Model}
\begin{Remark} See the classes of data poisoning attacks explored by Wang and Chaudhuri~\cite{wang2018data} against online learning, and of Chen et al~\cite{chen2017targeted}. for creating backdoors in deep-learning verification systems.\end{Remark}
\begin{Example} Financial market participants can (illegally) spoof by posting orders that will quickly be canceled in a ``momentum ignition'' strategy to lure others into buying or selling, as has been alleged to be occurring in high-frequency-trading~\cite{shorter2014high}. This differs from the earlier example in that the transactions are not bona-fide transactions which fool other agents, but are actually false evidence.\end{Example}
\begin{Example} Rating systems can be attacked by inputting false reviews into a system, or by discouraging reviews by those likely to be the least or most satisfied reviewers.\end{Example}
\begin{Model}{\textbf{4.2---Active Input Spoofing.}} As in (4.1), where the victim agent employs active learning. In this case, the opponent can potentially fool the system into collecting data that seems very useful to the victim from crafted poisoned sources.\end{Model}
\begin{Example} Honeypots can be placed, or Sybil attacks mounted by opponents to fool victims into learning from examples that systematically differ from the true distribution.\end{Example}
\begin{Example} Comments by users ``Max'' and ``Vincent DeBacco'' on Eisen's blog post about Amazon pricing suggested that it is very possible to abuse badly built linear pricing models on Amazon to receive discounts, if the algorithms choose prices based on other quoted prices~\cite{Eisen2011}.\end{Example}
\begin{Model}{\textbf{4.3---Input Filtering.}}  As in (4.1), but instead of generating false evidence, true evidence is hidden to systematically alter the distribution of events seen.\end{Model}
\begin{Example}Financial actors can filter the evidence available to other agents by performing transactions they do not want seen as private transactions or dark pool transactions.\end{Example}
\begin{Remark} There are classes of system where it is impossible to generate arbitrary false data points, but selective filtering can have similar effects.\end{Remark}
\begin{FailureMode} \textbf{Goal co-option} is when an opponent controls the system the Victim runs on, or relies on, and can therefore make changes to affect the victim's actions.\end{FailureMode}
\vspace{-15pt}\begin{Remark} Whenever the computer systems running AI and ML systems are themselves insecure, it presents a very tempting weak point that potentially requires much less effort than earlier methods of fooling the system.\end{Remark}
\begin{Model}{\textbf{5.1---External Reward Function Modification.}} Opponent $O$ directly modifies Victim $V$'s reward function to achieve a different objective than the one originally specified.\end{Model}
\begin{Remark} Slight changes in a reward function may have non-obvious impacts until after the system is~deployed.\end{Remark}
\begin{Model}{\textbf{5.2---Output Interception.}} Opponent $O$ intercepts and modifies Victim $V$'s output.\end{Model}
\begin{Model}{\textbf{5.3---Data or Label Interception.}} Opponent $O$ modifies externally stored scoring rules (labels) or data inputs provided to Victim $V$'s output.\end{Model}
\begin{Example}Xiao, Xiao, and Eckert explore a ``label flipping'' attack against support vector machines~\cite{xiaoadversarial} where modifying a limited number of labels used in the training set can cause performance to deteriorate severely.\end{Example}
\begin{Remark} As noted above, there are cases where generating false data may be impossible or easily detected. Modifying the inputs during training may create less obvious traces of an attack has occurred. Where this is impossible, access can also allow pure observation which, while not itself an attack, can allow an opponent to engage in various other exploits discussed earlier.\end{Remark}

To conclude the list of failure modes, it is useful to note a few areas where the failures are induced or amplified. This is when agents explicitly incentivize certain behaviors on the part of other agents, perhaps by providing payments. These public interactions and incentive payments are not fundamentally different from other failure modes, but can create or magnify any of the other modes. This is discussed in literature on the evolution of collusion, such as Dixon's treatment~\cite{Dixon2000}. Contra Dixon, however, the failure modes discussed here can prevent the collusion from being beneficial.
A~second, related case is when creating incentives where an agent fails to anticipate either the ways in which the other agents can achieve the incentivized target, or the systemic changes that are induced. These so-called ``Cobra effects''~\cite{Manheim2018b} can lead to both the simpler failures of the single-agent cases explored in Manheim and Garrabrant, and lead to the failures above. Lastly, as noted by Sandberg~\cite{Sandberg2018}, agents with different ``speeds'' (and, equivalently, processing power per unit time,) can exacerbate victimization, since older and slower systems are susceptible, and susceptibility to attacks only grows as new methods of exploitation are found.

\section{Discussion}

Multi-agent systems can naturally give rise to cooperation instead of competition,  as discussed in Leibo et al.'s 2017 paper~\cite{Leibo2017}.  The conditions under which there is exploitation rather than cooperation, however, are less well understood. A more recent paper by Leibo proposes that the competition dynamic can be used to encourage more complex models. This discusses coordination failures, but~the discussion of dynamics leading to the failures does not engage with the literature on safety or goal-alignment~{\cite{Leibo2019}}. Leibo's work, however, differs from most earlier work where multi-agent systems are trained together with a single goal, perforce leading to cooperative behavior, as in Lowe et al.'s heavily cited work, in which ``competitive'' dynamics are dealt with by pre-programming explicit models of other agent behaviors~\cite{Lowe2017}.

The failure modes outlined (accidental steering, coordination failures, adversarial misalignment, input spoofing or filtering, and goal co-option or direct hacking) are all due to models that do not fully account for other agent behavior. Because all models must simplify the systems they represent, the prerequisites for these failures are necessarily present in complex-enough systems where multiple non-coordinated agents interact. The problems of embedded agents discussed by Demski and Garrabrant~\cite{Demski2019} make it particularly clear that current approaches are fundamentally unable to fully represent these factors. For this and other reasons, mitigating the failures modes discussed here are not yet central to the work of building better ML or narrow AI systems. At the same time, some competitive domains such as finance are already experiencing some of these exploitative failures~\cite{shorter2014high}, and bots engaging in social network manipulation, or various forms of more direct interstate competition are likely engaging in similar strategies.

The failures seen so far are minimally disruptive. At the same time, many of the outlined failures are more problematic for agents with a higher degree of sophistication, so they should be expected not to lead to catastrophic failures given the types of fairly rudimentary agents currently being deployed. For this reason, specification gaming currently appears to be a mitigable problem, or as Stuart Russell claimed, be thought of as ``errors in specifying the objective, period'' \cite{krakovna2018russellcomment}. This might be taken to imply that these failures are avoidable, but the current trajectory of these systems means that the problems will inevitably worsen as they become more complex and more such systems are deployed, and the approaches used are fundamentally incapable of overcoming the obstacles discussed.

\subsection*{Potential Avenues for Mitigation}

Mitigations for these failures exist, but as long as the fundamental problems discussed by Demski and Garrabrant~\cite{Demski2019} are unaddressed, the dynamics driving these classes of failure seem unavoidable. Furthermore, such failures are likely to be surprising. They will emerge as multiple machine learning agents are deployed, and more sophisticated models will be more likely to trigger them. However, as~argued above, these failures are fundamental to interaction between complex agents. This means that while it is unclear how quickly such failures will emerge, or if they will be quickly recognized, it~is unquestionable that they will continue to occur. System designers and policymakers should expect that these problems will become intractable if deferred, and are therefore particularly critical to address now. It is be expected that any solution involves a combination of approaches~\cite{Sandberg2001}, though the brief overview of safety approaches below shows that not all general approaches to AI safety are helpful for multi-agent failures.

First, there are approaches that limit optimization. This can be done via satisficing, using approaches such as Taylor's Quantilizers, which pick actions at random from the top quantile of evaluated choices~\cite{Taylor2016}. Satisficing approaches can help in prevent exploiting other agents, or in preventing accidental overoptimization, but are not effective as a defense against exploitative agents or systemic failures due to agent interaction. Another approach limiting optimization is explicit safety guarantees. In extrema, this looks like an AI-Box, preventing any interaction of the AI with the wider world and hence preventing agent interaction completely. This is effective if such boxes are not escaped, but it is unclear if this is possible~\cite{Armstrong2012}. Less extreme versions of safety guarantees are sometimes possible, especially in domains where a formal model of safe behavior is possible, and~the system is sufficiently well understood. For example, Shalev-Shwartz et al. have such a model for self-driving cars, heavily relying on the fact that the physics involved with keeping cars from hitting one another, or other objects, is in effect perfectly understood~\cite{Shalev-Shwartz2017}. Expanding this to less well understood domains seems possible, but is problematic for reasons discussed elsewhere~\cite{Manheim2018f}.

Without limiting optimization explicitly, some approaches attempt to better define the goals, and~thereby reduce the extent of unanticipated behaviors.
These approaches involve some version of direct optimization safety. One promising direction for limiting the extent to which goal-directed optimization can be misdirected is to try to recognize actions rather than goals~\cite{Liu2016}. Human-in-the-loop oversight is another direction for minimizing surprise and ensuring alignment, though this is already infeasible in many systems~\cite{Danzig2018}. Neither approach is likely to be more effective than humans themselves are at preventing such exploitation. The primary forward-looking approach for safety is some version of ensuring that the goal is aligned, which is the bulk of what Yampolskiy and Fox refer to as AI safety engineering~\cite{Yampolskiy2013}.

In multi-agent contexts there is still a concern that because human values are complex,~\cite{Yudkowsky2011} exploitation is an intrinsically unavoidable pitfall in multi-agent systems. Paul Christiano's ``Distillation and Amplification'' approach involves safe amplification using coordinated multi-agent systems~\cite{christiano2018amplification}. This itself involves addressing some of the challenges with multi-agent approaches, and work on safe amplification using coordinated multi-agent systems in that context has begun~\cite{Irving2018}. In~that work, the coordinating agents are predictive instead of agentic, so the failure modes are more restricted. The methods suggested can also be extended to agentic systems, where they may prove more worrisome, and solving the challenges potentially involves mitigating several failure modes outlined here.

Between optimization-limiting approaches and AI safety engineering, it is possible that many of the multi-agent failures discussed in the paper can be mitigated, though not eliminated.  In addition, there will always be pressure to prioritize performance as opposed to safety, and safe systems are unlikely to perform as quickly as unsafe ones~\cite{Danzig2018}. Even if the tradeoff resolves in favor of slower, safer systems, such systems can only be created if these approaches are further explored and the many challenges involved are solved before widespread deployment of unsafe ML and AI. Once the systems are deployed, it seems infeasible that safer approaches could stop failures due to exploiting and exploitable systems, short of recalling them. This is not a concern for the far-off future where misaligned superintelligent AI poses an existential risk. It is instead a present problem, and it is growing more serious along with the growth of research that does not address it.

\section{Conclusions: Model Failures and Policy Failures}

Work addressing the failure modes outlined in the paper is potentially very valuable, in part because these failure modes are mitigable or avoidable if anticipated. AI and ML system designers and users should expect that many currently successful but naive agents will be exploited in the future. Because of this, the failure modes are likely to become more difficult to address if deferred, and are therefore particularly critical to understand and address them preemptively. This may take the form of systemic changes such as redesigned financial market structures, or may involve ensuring that agents have built-in failsafes, or that they fail gracefully when exploited.

At present, it seems unlikely that large enough and detected failures will be sufficient to slow the deployment of these systems. It is possible that governmental actors, policymakers, and commercial entities will recognize the tremendous complexities of multiparty coordination among autonomous agents and address these failure modes, or slow deployment and work towards addressing these problems even before they become catastrophic. Alternatively, it is possible these challenges will become apparent via limited catastrophes that are so blatant that AI safety will be prioritized. This depends on how critical the failures are, how clearly they can be diagnosed, and whether the public demands they be addressed.

Even if AI amplification remains wholly infeasible, humanity is already deploying autonomous systems with little regards to safety. The depth of complexity is significant but limited in current systems, and the strategic interactions of autonomous systems are therefore even more limited. However, just as AI for poker eventually became capable enough to understand multi-player interaction and engage in strategic play, AI in other systems should expect to be confronted with these challenges. We do not know when the card sharks will show up, or the extent to which they will make the games they play unsafe for others, but we should admit now that we are as-yet unprepared for~them.





\vspace{6pt}


\funding{This research was funded in large part by a grant from the Berkeley Existential Risk Initiative.}

\acknowledgments{I would like to thank a subset of the anonymous reviewers in both the first and second submission for very helpful comments, and thank Roman Yampolskiy for encouraging me to write and revise the paper, despite setbacks.}


\conflictofinterests{The author declares no conflict of interest. The funders had no role in the writing of the manuscript, nor in the decision to publish the results.}

\reftitle{References}







\end{document}